# PREFALLKD: PRE-IMPACT FALL DETECTION VIA CNN-VIT KNOWLEDGE DISTILLATION

*Tin-Han Chi[1], Kai-Chun Liu[2], Chia-Yeh Hsieh[3], Yu Tsao[2], Chia-Tai Chan[1]*

[1]Department of Biomedical Engineering, National Yang Ming Chiao Tung University, Taiwan
[2]Research Center for Information Technology Innovation, Academia Sinica, Taiwan
[3]Bachelor's Program in Medical Informatics and Innovative Applications, Fu Jen Catholic University, Taiwan

## ABSTRACT

Fall accidents are critical issues in an aging and aged society. Recently, many researchers developed "pre-impact fall detection systems" using deep learning to support wearable-based fall protection systems for preventing severe injuries. However, most works only employed simple neural network models instead of complex models considering the usability in resource-constrained mobile devices and strict latency requirements. In this work, we propose a novel pre-impact fall detection via CNN-ViT knowledge distillation, namely PreFallKD, to strike a balance between detection performance and computational complexity. The proposed PreFallKD transfers the detection knowledge from the pre-trained teacher model (vision transformer) to the student model (lightweight convolutional neural networks). Additionally, we apply data augmentation techniques to tackle issues of data imbalance. We conduct the experiment on the KFall public dataset and compare PreFallKD with other state-of-the-art models. The experiment results show that PreFallKD could boost the student model during the testing phase and achieves reliable F1-score (92.66%) and lead time (551.3 ms).

## 1. INTRODUCTION

World Health Organization (WHO) has pointed out that falls are the second leading cause of unintentional injury deaths worldwide [1]. Fall can also cause post-fall syndrome (e.g., the fear of falling again) which may reduce the independence of their daily life [2]. Currently, many researchers developed "pre-impact fall detection systems" using wearable sensors [3] to support fall protection systems for preventing severe injuries [4]. The purpose of pre-impact fall detection systems is to detect the occurrence of falls during falling with a lead time which is determined as the period between the fall detection of the system and the ground impact. A longer lead time indicates that the protection system has more response time to activate the fall protection device as the duration of falling is less than 800ms in most fall instances [5]. It shows the requirement to develop reliable pre-impact fall detection systems to quickly and accurately detect early falls.

For more robust detection performance, many researchers toward to apply deep learning models to backbone of pre-impact fall detection systems. Convolutional neural network (CNN) was generally applied in fall event detection systems, which can extract spatial features of raw data with convolutional filters. Li et al. [6] firstly proposed an approach for pre-impact fall detection during gait rehabilitation training based on a 3D CNN using RGB images, which achieved 100% accuracy from 5 trainees. Yu et al. [7] applied the convolutional long short-term memory (ConvLSTM) model to extract spatial-temporal features to perform three-class classification on non-fall, pre-impact fall, and fall. Their state-of-the-art (SOTA) model achieved accuracy of 93.22%, 94.48%, and 98.66% for non-fall, pre-impact fall, and fall classes, respectively. However, most works only employed simple neural network models instead of complex models to ensure usability in resource-constrained mobile devices and strict latency requirements. Such concern may limit the performance of the detection model.

To overcome the issues, we develop a pre-impact fall detection system using knowledge distillation (KD) [8-10], namely PreFallKD. By this approach, we could transfer the detection knowledge from the pre-trained larger model (teacher) to the smaller model (student). It allows the system to keep the computing power efficient and boost the detection performance using the student model. In this study, we selected typical CNN as the student model and vision transformer (ViT) [11] as the teacher model, while the effectiveness of ViT has been validated on different classification tasks, including speech recognition [12] and image classification [13]. Moreover, we employed data augmentation, including random gaussian noise and random magnitude scale [14], to increase the size of the minority class and total number for more robust model training.

The main contributions of this work are summarized as follows: (1) To the best of our knowledge, this is the first pre-impact fall detection conducting the KD method. (2) We conducted the experiment on the KFall public dataset and compare PreFallKD with other SOTA models. (3) The experiment results show that the proposed PreFallKD could outperform SOTA models with longer leading time.

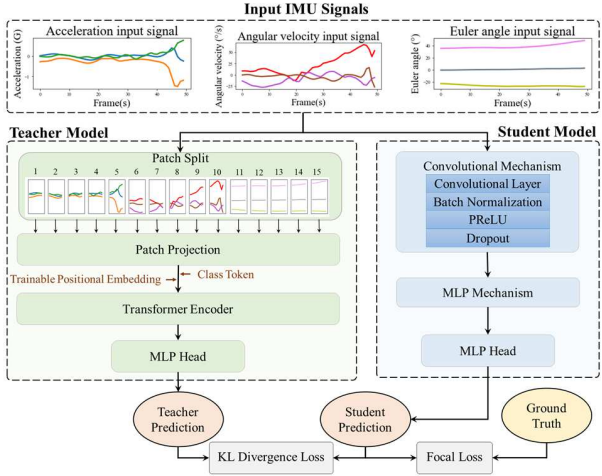

**Fig. 1**: An overview of proposed PreFallKD framework.

## 2. THE PROPOSED METHOD

The overview of PreFallKD is shown in Fig. 1. The input window is the 50 frames IMU signals, which include triaxial acceleration data, triaxial angular velocity data, and triaxial Euler angle data. The ViT-tiny is the teacher model and the lightweight CNN is the student model. The student model can learn the high dimension knowledge from the teacher model by Kullback-Leibler divergence loss function (KL Divergence Loss) and learn ground truth by Focal loss.

### 2.1. Data augmentation

To decrease the negative effect of imbalance problems, we oversampled all the pre-impact fall windows 6 times to balance the number of windows between the ADL windows and pre-impact fall windows in the training dataset. Then, we applied two data augmentation techniques to increase the total number of training data. One is to add the random gaussian noise to each window, as shown in Equation (1) [14], where $D_\epsilon$ is the data mixed with the gaussian noise, $\sigma$ is the standard deviation of each axis, $N$ is the random normal distribution function with a range from 0 to 1, $s$ is the strength parameter and $D$ is the original data. The other is to adjust the magnitude scale of windows, as shown in Equation (2), where $D_\mu$ is the data with magnitude scale, $U$ is the random discrete uniform distribution function with a range from 0.75 to 1.25, and $s$ is the strength parameter. We determine $s$ as -0.5.

$$D_\epsilon = 0.25 \times \sigma \times N(0,1) \times sigmoid(s) + D \quad (1)$$

$$D_\mu = U(0.75, 1.25) \times sigmoid(s) \times D \quad (2)$$

### 2.2. Teacher model

The ViT framework is shown as Fig. 1. We reshape input windows from $x \in R^{C*L*A}$ to $x \in R^{N*(C*L_p*A_p)}$ to fit the input sequences of ViT, where $C$ is the number of channels, $L$ is the window length, $A$ is the number of window axis, $N$ is the number of window patches, $L_p$ and $A_p$ are the length and axis of each patch. Then, these patches are projected to patch vectors by a linear projection. The patch vectors provide detailed time-series characteristics. Subsequently, we use the trainable class token and position embeddings to strengthen the spatial information which were used in the original ViT and Bidirectional Encoder Representations from Transformers (BERT) [15]. Finally, these embedding feature maps are the inputs to the transformer encoder, which contains multi-head attention, layer normalization, and multilayer perceptron (MLP) mechanisms. Considering the number of IMU data in this work is far less than the number of images in the original ViT [11], we build a "tiny" ViT model (ViT-tiny) with 3 layers, 3 heads, 64 hidden size, 256 MLP size, dropout 0.2, and $L_p$ is 10, $A_p$ is 3.

### 2.3. Student model

The lightweight convolutional neural network (CNN) is composed of two convolutional mechanisms and one MLP mechanism. Each convolutional mechanism is composed of 64 3-by-3 filters, a batch normalization, a max pooling layer with a 1-by-2 filter, and a dropout operation of 0.1. We use the parametric rectified linear unit (PReLU) [16] as an activation function, as shown in Equation (3), where the coefficient α of the negative part in PReLU is adaptively learned. It enhances model fitting with nearly zero extra computational cost and little overfitting risk compared to ReLU. The MLP mechanism is composed of two linear projection layers with a PReLU and a log SoftMax activation function.

$$PReLU(x) = \begin{cases} \alpha x, & x < 0 \\ x, & x \geq 0 \end{cases} \quad (3)$$

### 2.4. Knowledge Distillation (KD)

Knowledge Distillation [8] allows PreFallKD to train a small-size model to distill the knowledge of the larger-size model. The soft labels from the teacher model provide more information than the hard labels for the student model. Because KD can improve the performance of the small model, we can deploy it on devices that are limited by low power or resources.

The loss function of PreFallKD to train student model is defined as Equation (4), where $\beta$ is the automatically adjust parameter which controls the importance of the teacher model's knowledge, $L_F$ is the Focal loss function [17], $L_K$ is the KL divergence loss, $P_s$ is the predicted probability of student network and $P_t$ is the predicted probability of teacher network. We initial $\beta$ as 1 and it would drop with each epoch. We expect the student can learn more detection knowledge from the teacher model at the beginning and gradually reduce its dependence on the teacher.

The Focal Loss function, $L_F$, is defined as Equation (5). It used an adjustable term for cross-entropy loss. The value of $\alpha$ is set to control the shared weight of positive and negative events. The instances with more samples would be set at a smaller $\alpha$ to reduce the weight. The value of $\gamma$ is the scaling factor used to focus on the hard events. We implement $\alpha = 0.25$ and $\gamma = 2$. The KL divergence loss function is defined as Equation (6), where $m$ is the number of windows and $T$ is the temperature parameter related to the contribution of the teacher model. $T$ is set to 1.

$$L_{KD} = (1-\beta) \times L_F(P_s) + \beta \times L_K(P_s, P_t) \quad (4)$$

$$L_F(P_s) = -\alpha(1-P_s)^\gamma \log(P_s) \quad (5)$$

$$L_K = \frac{1}{m} \times \sum_{k=0}^{m} \frac{P_t}{T} \times \left(\log \frac{\frac{P_t}{T}}{\frac{P_s}{T}}\right) \quad (6)$$

## 3. EXPERIMENTAL

### 3.1. Public dataset

KFall is the first public dataset suitable for pre-impact fall detection, not just for post-fall detection. This dataset was collected from 32 healthy young Korean males (age: 24.9 ± 3.7 years; height: 174.0 ± 6.3 cm; weight: 69.3 ± 9.5 kg). It incorporated nine-axis inertial sensor data and temporal labels. A total of 21 types of activities of daily living (ADLs) and 15 types of simulated falls were included. KFall contains a total of 5,075 instances, including 2,729 ADL and 2,346 fall instances. Each instance is composed of triaxial acceleration data, triaxial gyroscope data, and triaxial Euler angle data. The inertial sensor was configured at a frequency of 100 Hz and attached to the low back of each subject. The temporal labels in each fall instance contain the fall onset moment and the fall impact moment based on synchronized motion videos. The fall onset moment is defined as the moment the fall begins; the fall impact moment is the moment the body hit the ground. Detailed information is introduced in [18].

### 3.2. Data preparation

For ADL instances preprocessing, each ADL instance is divided into ten parts equally. Then, we randomly captured a window with a size of 50 frames (0.5s) from each part. The windows obtained from ADL instances are labeled as ADL windows. This way is to reduce the imbalance problem of ADL windows and pre-impact fall windows, but catch the entire instance's characteristics.

For fall instance preprocessing, the sliding window technique was applied with the size of 50 frames and a sliding step of 10 frames (0.1s) to the time interval from the starting point to the fall impact moment. The windows before the fall onset moment were labeled as ADL windows. The windows reaching or over the fall onset moment for more than 5 frames (0.05s) were labeled as pre-impact fall windows.

During the training phase, all ADL windows and pre-impact fall windows would feed to PreFallKD. During the testing phase, the first 3 pre-impact windows reaching over the fall onset moment in each fall instance were collected and all ADL windows were involved in the testing dataset because the period of falling is less than 800ms in most fall instances [19]. Such testing data could simulate real-world situations to test the detection reliability and lead time. Overall, the training phase has 270657 ADL and 290484 pre-impact windows, and the testing phase has 46826 ADL and 7023 pre-impact windows.

### 3.3. Performance evaluation

This work utilizes leave one group out cross validation approach to evaluate performances. The approach divides all subjects into 5 groups and iterates 5 times until each group was used as the testing set. For each iteration, one of the group's data in the testing dataset is for testing, two of the groups' data in the training dataset is for teacher pre-training, and the remaining two groups' data in the training dataset is for student-training. This method ensures that the model does not learn any characteristics about the subjects in the testing dataset in advance.

Five common evaluation metrics are considered in this work to assess the models, including accuracy, recall, precision, specificity, and F1-score, where true positive (TP), false positive (FP), true negative (TN) and false negative (FN) are defined as follows:
- TP: A window is determined as TP if at least one window labeled with "fall" is identified as fall events.
- TN: A window is determined as TN if the window labeled with "ADL" is identified as ADLs.
- FP: A window is determined as FP if the window labeled with "ADL" is identified as fall events.
- FN: A window is determined as FN if all windows labeled with "fall" are identified as ADLs.

In addition to these metrics, lead time, model parameters, and computation time are used to evaluate the model's computing performance. The longer lead time indicates that the protection system has more response time. To estimate the computation time in the real world, the simulation environment comprises the STM32L476JGY micro control units (MCU) with an operating frequency of 80 MHz. The $MCU_{Flops}$ value can be estimated as 11.4 MFLOPS [20].

### 3.4. Implementation detail

In the teacher pre-training and student-training process, the AdamW optimizer [21] is utilized. The cosine learning rate schedule with warmup is applied with a learning rate of 0.001 and 10 warmup epochs. The number of waves in the cosine schedule is 0.5. The batch size is 64. Specifically, the teacher model is trained for 150 epochs and the student model is trained for 100 epochs. The teacher model only uses the Focal Loss function. The student model uses our KD loss function. Our proposed models are implemented on python 3.8 in a Window 11 environment. The models were trained and tested on a computer equipped with an Intel core i7-12700F CPU, 32GB RAM, and a Nvidia GTX 3080 GPU.

Table 1: Performance of different models evaluated on the KFall dataset

| Model | Preprocessing | KD | Accuracy (%) | Precision (%) | Recall (%) | Specificity (%) | F1-score (%) |
|---|---|---|---|---|---|---|---|
| Baseline-CNN | No | No | 96.56 | **92.36** | 80.27 | 99.01 | 85.89 |
| Lightweight-CNN | Yes | No | 97.33 | 84.97 | **96.58** | 97.44 | 90.40 |
| CNNLSTM | Yes | No | 97.67 | 88.35 | 94.58 | 98.13 | 91.36 |
| ViT-tiny | Yes | No | **98.36** | 92.02 | 95.73 | **99.36** | **93.84** |
| PreFallKD | Yes | Yes | 98.05 | 90.62 | 94.79 | 98.53 | 92.66 |

Table 2: The computing performance of different models evaluated on the KFall dataset

| Model | Lead time (ms) | Parameter (byte) | Computation time (ms) |
|---|---|---|---|
| Baseline-CNN | 511.2±79 | **59557** | **36.8** |
| Lightweight-CNN | 537.2±71 | **59557** | **36.8** |
| CNNLSTM | 493.5±62 | 81093 | 98.5 |
| ViT-tiny | 235.4±62 | 251010 | 348.6 |
| PreFallKD | **551.3±66** | **59557** | **36.8** |

## 4. EXPERIMENTAL RESULTS

To demonstrate the superiority of the proposed framework for pre-impact fall detection, we implement several SOTA models on the same dataset. The experimental setup of all models is identical to the proposed method, and their parameters are optimized. The models are as follows:

- **Baseline-CNN** is composed of 2 convolutional blocks and 2 fully connected layers. The training process excludes any data augmentation and KD processes. This model is used to understand the effect of our data preprocessing.
- **Lightweight CNN** is the structure of our student model.
- **CNN-LSTM** is modified from the SOTA model in pre-impact fall detection [18], which is composed of 2 convolutional mechanisms and 2 LSTM mechanisms.
- **ViT-tiny** is the same as our teacher model.

Except the baseline-CNN, the training data are processed with data augmentation techniques.

Table 1 demonstrates the performance of different models on the KFall dataset. Pre-impact fall detection system using ViT-tiny achieves the best detection performance with 98.36% accuracy and 93.84% F1-score, which outperforms SOTA CNNLSTM [7]. Such results prove the effectiveness of the self-attention mechanism in pre-impact fall detection systems compared to other typical neural network models. Moreover, the experiments present that the proposed PreFallKD using knowledge distillation via CNN-ViT could comprehensively boost the detection performance of student model (lightweight CNN). It improves 0.72% accuracy, 5.65% precision, 1.09% specificity and 2.26% F1-score. This result shows the KD method successfully transfers the high dimension knowledge and the long-term dependencies learned from ViT-tiny to lightweight CNN. Interestingly, the performance of PreFallKD outperforms SOTA model. Table 1 also shows that baseline-CNN got the worst accuracy (96.56%), recall (80.26%), and F1-score (85.89%). It often misclassified the pre-impact fall as ADL. Compared to Baseline-CNN, light weight CNN using data augmentation approaches achieves the better 96.58% recall and 90.40% F1-score. The main reason is that the smaller number of pre-impact fall windows hinder the detection in learning effective features to distinguish pre-fall impact from ADL. Our experiments display that efficient data augmentation techniques could help pre-impact fall detection systems tackle the imbalance issue and achieve better detection performance.

Table 2 demonstrates the computational performance of different models in terms of lead time, model parameters, and computation time of the model. Baseline-CNN, lightweight-CNN, and student model of PreFallKD share the same model architecture, so their model parameters and computation time are identical. Obviously, the lead time of ViT is 235.4±62 notably less than that of other models since it requires larger computing time and parameters compared to other models. Although we greatly reduced the number of layers and hidden size of ViT, its computation efficiency may fail to support the fall protection system requiring 333ms to inflate the protection airbag [22]. Our results show that PreFallKD has the longest lead time of 551.3±66 ms which is sufficient to activate protective devices in time. It proves the superiority of PreFallKD considering the balance between detection performance and response time.

## 5. CONCLUSIONS

We develop a pre-impact fall detection system via CNN-ViT knowledge distillation, namely PreFallKD, to boost detection performance and reduce computational loading during the testing phase. Additionally, we validate the effectiveness of data augmentation techniques in pre-impact fall detection systems while tackling issues of data imbalance. The results show that PreFallKD achieves reliable F1-score (92.66%) and lead time (551.3 ms). In future work, we plan to combine other model compression techniques (e.g., punning and quantization) to improve the system performance.